\documentclass[prl, twocolumn, amsmath,amssymb,showpacs]{revtex4-1}

\usepackage{graphicx}% Include figure files
\usepackage{dcolumn}% Align table columns on decimal point
\usepackage{bm}% bold math

\begin{document}
%================= Title, affiliations etc =============================================%
\title{Probing the interiors of the ice giants: Shock compression of water to 700 GPa and 3.8 g/cm$^3$}
\author{M.D. Knudson}
\author{M.P.\ Desjarlais}
\author{R.W. Lemke}
\author{T.R. Mattsson}
\affiliation{Sandia National Laboratories, Albuquerque, New Mexico 87185, USA.}
\author{M. French}
\author{N. Nettelmann}
\author{R. Redmer}
\affiliation{Universit{\"a}t Rostock, Institut f{\"u}r Physik, D-18051 Rostock, Germany.}

\date{\today}

%============================= ABSTRACT ====================================%
\begin{abstract}
Recently there has been tremendous increase in the number of identified extra-solar planetary systems. Our understanding of their formation is tied to exoplanet internal structure models, which rely upon equations of state of light elements and compounds like water. Here we present shock compression data for water with unprecedented accuracy that shows water equations of state commonly used in planetary modeling significantly overestimate the compressibility at conditions relevant to planetary interiors. Furthermore, we show its behavior at these conditions, including reflectivity and isentropic response, is well described by a recent first-principles based equation of state. These findings advocate this water model be used as the standard for modeling Neptune, Uranus, and ``hot Neptune'' exoplanets, and should improve our understanding of these types of planets.
\end{abstract}


\maketitle

The past several years have seen a virtual explosion in the number of extra-solar planets discovered. Two rapidly growing populations of exoplanets are ice giants referred to as ``hot Neptunes'' and ``mini-Neptunes'';~\cite{Borucki} planets roughly the same size as or, respectively, smaller than Neptune and Uranus that transit their host stars at significantly smaller radii, resulting in higher temperatures than the ice giants in our solar system. Understanding of the composition and formation of these planets, and thus development of these planetary systems, relies on our knowledge of the equation of state (EOS) of light elements and compounds like water, over a wide pressure and temperature range. To date much of the modeling of ice giants has employed the ANEOS~\cite{ANEOS} and Sesame~\cite{Sesame} models for water that were developed decades ago~\cite{Grasset,Swift}. Discrepancies between these EOS models lead to significant differences in predicted radius evolution of Neptune-mass planets. Depending upon the total amount of heavy elements, and their distribution within the planetary interior, the resulting variation in predicted radius at a given age due to the water EOS can range between 5 and 30\% ~\cite{Chabrier}. This is a major factor in preventing accurate determination of exoplanet internal composition from their observed radius.

Recent quantum molecular dynamics (QMD) calculations of water~\cite{FrenchPRB2009,FrenchJPCM} suggest an EOS that differs significantly from ANEOS and Sesame. Notably, when incorporated into planetary models, this first-principles (FP) based EOS predicts a $\sim$20\% cooler core temperature for Neptune and Uranus~\cite{Fortney}. The conductivity properties of this FP model are also noteworthy~\cite{FrenchPRB2010}, suggesting that water is super-ionic~\cite{Cavazzoni,Mattsson12} at high densities, $\rho$, and low temperatures, $T$, relevant to planets such as Uranus and Neptune. This predicted property plays a key role in dynamo models to explain the enigmatic magnetic field structure of these planets~\cite{Stanley,Redmer}. Another important result is derived from the predicted phase diagram of water: the icy giants Uranus and Neptune perhaps contain no ``ice'' but dissociated water at a high ionic conductivity, even less so would close-in exoplanets. Hot and mini-Neptunes may even comprise water plasma with substantial electronic conduction. However, the FP EOS for water has not been widely accepted due to its inability to reproduce results from laser driven shock wave experiments in the Mbar regime~\cite{Celliers}.

We present results of magnetically accelerated flyer-plate experiments on water performed at the Sandia Z machine~\cite{MatzenPoP2005}, a pulsed power accelerator capable of producing extremely large current ($\sim$20 MA) and magnetic field densities ($\sim$10 MG) within a short circuit load. These data, in the range of 100-450 GPa along the Hugoniot -- the locus of end states achievable through compression by large amplitude shock waves -- have considerably higher precision than data obtained with previously used methods, and support the FP EOS for water. The high precision stems from the ability to perform well-defined flyer-plate experiments on Z; the magnetic pressure (\textgreater 500 GPa) can propel the outer anode to velocities approaching 30 km/s, enabling high-precision, plate-impact EOS measurements in the TPa regime~\cite{Lemke,Science,Quartz}. Furthermore, and more significantly, the present work obtained re-shock data of water in the range of 200-700 GPa. These data, at high $\rho$ and low $T$, provide a stringent test of the isentropic response of water in the several Mbar regime, which is directly relevant to the conditions of interest for planetary modeling of Neptune, Uranus~\cite{Fortney,Redmer}, and presumably water-rich exoplanets such as the hot Neptune GJ436b~\cite{Fortney,Nettelmann2011,Nettelmann2010}. Finally, reflectivity on the Hugoniot was measured and compared to FP calculations for water~\cite{FrenchPoP}.

An aluminum flyer-plate~\cite{SM} was magnetically accelerated to peak velocities of 12-27 km/s across a 3-4 mm vacuum gap~\cite{Lemke}. The flyer-plate velocity was monitored throughout the entire trajectory using a Velocity Interferometer System for Any Reflector (VISAR~\cite{Barker}), at locations above and below an aluminum water cell~\cite{SM}. A rear quartz window in the cell provided optical access to the sample. In some cases an additional quartz plate was placed between the aluminum drive plate and the water sample, enabling data to be obtained using two different materials as the high-pressure standard, thereby increasing confidence in the measurements. Impact with the cell generated a strong, multi-Mbar shock wave in the aluminum drive plate. This shock was then transmitted either directly into the water sample, or into a quartz plate and then into the water sample. Upon reaching the rear quartz window, the shock was transmitted into the window and reflected back into the water,which re-shocked the water to a higher $P$ and $\rho$. In all cases the shock waves in the water and quartz were of sufficient amplitude that the resulting shocked material was reflecting~\cite{Celliers,Quartz,Hicks}, enabling the shock velocities to be directly measured using the VISAR. A total of 18 diagnostic channels were utilized for each experiment, enabling multiple, redundant measurements to be made, resulting in an overall uncertainty in the measured flyer-plate and shock velocities of a few tenths of a percent~\cite{SM}.

\begin{figure}
\includegraphics*[width=7cm]{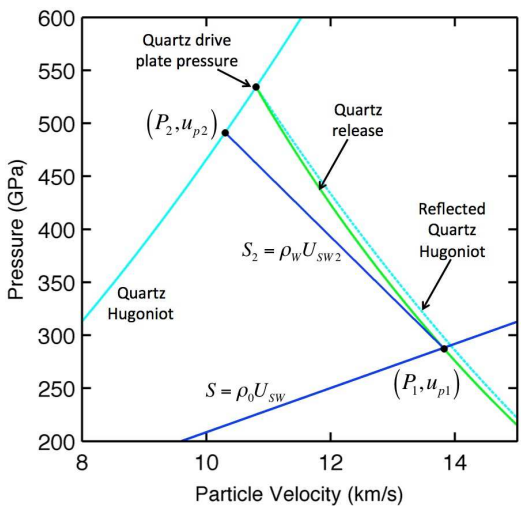}
\caption{$P-u_p$ diagram for the water experiments for the case with the additional quartz drive plate.}
\label{fig:match}
\end{figure}

The shocked state of the water was determined using the impedance matching technique and the Rankine-Hugoniot (RH) jump relations~\cite{Duvall}, a set of conditions derived by considering conservation of mass, momentum, and energy across a steady propagating shock wave. The shocked state of the aluminum (quartz) drive plate was determined from the known Hugoniot of aluminum~\cite{aluminum} (quartz~\cite{Quartz}) and the measured flyer-plate (quartz shock) velocity; this defined a point in the pressure - particle velocity ($P-u_p$) plane, as shown in Fig.~\ref{fig:match}. When the shock transits into the water, a release wave propagates back toward the flyer-plate, and thus the state of the drive plate is constrained to lie on a release adiabat from this point in the $P-u_p$ plane, shown in Fig.~\ref{fig:match} as the green line. The shocked state of the water is constrained to lie along a chord in the $P-u_p$ plane with slope given by the product of the measured shock velocity of water, $U_{sw}$, and the known initial density. The intersection of these two curves provides $P$ and $u_p$, shown in Fig.~\ref{fig:match} as $(P_1,u_{p1})$; The RH jump relations then provide $\rho$ in the shocked state. Uncertainties in all kinematic values were determined through a Monte Carlo technique, which uses a statistical process for propagation of all random measurement errors and systematic errors in the standards~\cite{SM}. Using this technique, the one-sigma uncertainties in $P$ and $\rho$ were found to be ~0.5\% and ~1\%, respectively.

\begin{figure}
\includegraphics*[width=8cm]{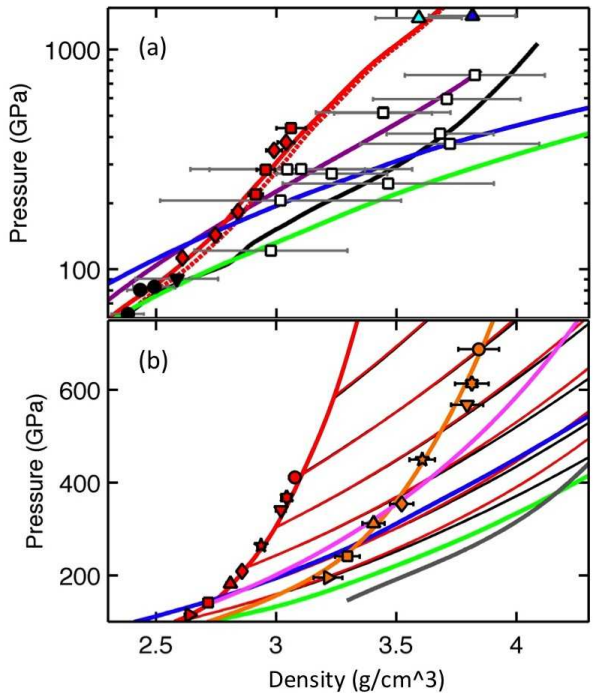}
\caption{(a) Water $P-\rho$ Hugoniot. \textit{Models}: magenta line, ANEOS~\cite{ANEOS}; black line, Sesame~\cite{Sesame}; dotted red line, FP~\cite{FrenchPRB2009}; solid red line, FP~\cite{FrenchJPCM}. \textit{Data}: red diamonds (squares), this work using aluminum (quartz) standard; open squares, Celliers \textit{et al.}~\cite{Celliers}; black circles, Mitchell and Nellis~\cite{Mitchell}; black triangle, Volkov \textit{et al.}~\cite{Volkov}; blue (cyan) triangle, Podurets \textit{et al.}~\cite{Podurets}, as reported (reanalyzed~\cite{SM}). \textit{Planetary adiabats}: green line, Neptune~\cite{Fortney} (Uranus similar); blue line, GJ436b~\cite{Nettelmann2011,Nettelmann2010,SM}. (b) Water double-shock Hugoniot. \textit{Models}: Thick (thin) red lines, FP first shock (re-shock) Hugoniots~\cite{FrenchPRB2009}; black lines, FP isentropes~\cite{FrenchPRB2009}; (orange, pink, gray) lines, (FP~\cite{FrenchPRB2009}, ANEOS~\cite{ANEOS}, Sesame~\cite{Sesame}) double-shock envelopes. \textit{Data}: red (orange) symbols, shock (re-shock) Hugoniot data, this work. \textit{Planetary adiabats}: as in panel (a).}
\label{fig:hug}
\end{figure}

A total of 8 Hugoniot experiments were performed over the range of 100 to 450 GPa. Results of these experiments are shown as the red symbols in Fig.~\ref{fig:hug}(a). Also shown are Hugoniot data of Mitchell and Nellis~\cite{Mitchell}, Volkov \textit{et al.}~\cite{Volkov}, Celliers \textit{et al.}~\cite{Celliers}, and Podurets \textit{et al.}~\cite{Podurets}, and the predicted Hugoniot response from ANEOS~\cite{ANEOS}, Sesame 7150~\cite{Sesame}, and the recent FP EOS model of French \textit{et al.}~\cite{FrenchPRB2009,FrenchJPCM}. Note that a reanalysis of the nuclear driven datum of Podurets \textit{et al.}, using an improved aluminum standard for impedance matching~\cite{aluminum}, resulted in a slight decrease in $\rho$. The low-$P$ end of our data is in good agreement with the gas gun data of Mitchell and Nellis and the explosively driven shock data of Volkov \textit{et al.} In contrast, our data are significantly less compressible than the laser driven data of Celliers \textit{et al.}, which tend to support the much more compressible ANEOS and Sesame Hugoniots, albeit with significantly large uncertainty and scatter. The vastly reduced uncertainty in $\rho$ for our data, roughly an order of magnitude, strongly suggest that water is much less compressible than the ANEOS and Sesame models predict, and that water is instead very accurately described by the FP EOS of French \textit{et al.} Furthermore, the reanalyzed Podurets \textit{et al.} datum is also in very good agreement with the FP EOS. Thus, with the exception of the Celliers \textit{et al.} data, the FP based model for water matches all experimental Hugoniot data up to ~1.4 TPa.

In all 8 of the Hugoniot experiments described above, the reflected shock from the rear quartz window drove the water from the Hugoniot state to a re-shocked state at higher $P$ and $\rho$. The measured shock velocity in the water immediately prior to reflection from the rear quartz window defined the initial shocked state of the water. The measured shock velocity in the rear quartz window and the known Hugoniot of quartz provided the double-shocked $P$ and $u_p$ for water, shown in Fig.~\ref{fig:match} as $(P_2,u_{p2})$. The velocity of the second shock in the water, $U_{sw2}$, was then determined by the RH jump relations using the change in $P$, $(P_2-P_1 )$, and $u_p$, $(u_{p2}-u_{p1} )$. The re-shock $\rho$ was then determined from $U_{sw2}$, the first shock $\rho$, and $(u_{p2}-u_{p1})$. Using the Monte Carlo technique, the one-sigma uncertainties in $P$ and $\rho$ for the re-shock states were found to be ~0.5-1\% and ~1-2\%, respectively. Although the uncertainty for the re-shock data is larger than that for the principal Hugoniot data (entirely due to the larger uncertainty in the initial state), the accuracy of the present data is a significant improvement over previous re-shock data of Mitchell and Nellis~\cite{Mitchell} (uncertainty in $\rho$ of 4-14\%) and the pre-compressed Hugoniot data of Lee \textit{et al.}~\cite{Lee}, (uncertainty in $\rho$ of 5-10\%).

The re-shock data for water are shown in Fig.~\ref{fig:hug}(b), where first and second shock states are correlated by like symbols. Also shown are several FP re-shock Hugoniots (thin red lines) and isentropes (thin black lines) for comparison~\cite{FrenchPRB2009}. These re-shock Hugoniots along with the known Hugoniot of quartz~\cite{Quartz} were used to determine the double-shock envelopes –- the locus of end states achievable through shock and re-shock using a quartz anvil: FP (orange line), ANEOS~\cite{ANEOS} (pink line), and Sesame~\cite{Sesame} (gray line). These re-shock data further confirm the less compressible response of water above ~100 GPa.

Note that the FP re-shock Hugoniots (red) and isentropes (black) are nearly coincident over the $\rho$ range accessed through the re-shock experiments. This is due to a second order contact for the Hugoniot and isentrope at the initial state~\cite{Duvall}, which is most easily seen by expanding the entropy as a function of volume in a Taylor series. This implies that the Hugoniot and isentrope are very close in $P$ and $\rho$ until, at large compression, the rise in $T$ associated with the irreversible shock becomes large enough that thermal pressures become significant. In the range investigated in this study, the difference in $T$ between the re-shock Hugoniot states and the isentrope at the re-shock $\rho$, as determined by the FP EOS~\cite{FrenchPRB2009}, ranged from 200K (out of 6800K) to 330K (out of 40000K) at the lowest and highest $P$, respectively. This makes such a re-shock measurement the best possible test of the isentropic response of the EOS model in this range of $P$ and $\rho$. Thus the present data validates the isentropic response of the FP EOS in the $P$ and $\rho$ regime that is intersected by the water-rich models of Neptune and Uranus~\cite{Fortney,Redmer}, shown in green, and the exoplanet GJ436b~\cite{Nettelmann2011,Nettelmann2010,SM}, shown in blue.

The VISAR was also used to infer reflectivity, $R$, of water (at 532 nm) along the Hugoniot. A quadrature VISAR was used for all experiments, which provides four measures of the interference signal at $90\,^{\circ}$ intervals. The signals at $180\,^{\circ}$ intervals can be subtracted, ensuring the remaining signal only includes coherent reflected laser light (incoherent light, such as self-emission from the hot plasma, would equally contribute to all four quadrature signals). Comparison of the magnitude of these subtracted signals before and after shock breakout from the water to the quartz rear window provides a relative measure of the shocked water $R$ with respect to shocked quartz~\cite{Hicks}. The uncertainty in $R$ was taken to be the linear sum of the standard deviation of the inferred $R$ from the nine independent VISAR signals obtained from each water cell and the reported uncertainty in $R$ of shocked quartz~\cite{Hicks}.

\begin{figure}
\includegraphics*[width=6.5cm]{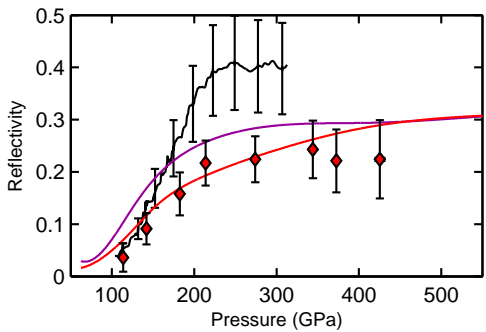}
\caption{Reflectivity (532 nm) along the principal Hugoniot of water. \textit{Models}: red (magenta) line, FP calculations with HSE (PBE) functionals~\cite{FrenchPoP}. \textit{Data}: black line, Celliers \textit{et al.}~\cite{Celliers}; red diamonds, this work.}
\label{fig:reflectivity}
\end{figure}

$R$ data along the Hugoniot are shown in Fig.~\ref{fig:reflectivity}. Also shown are data from Celliers \textit{et al.}~\cite{Celliers} and the predicted $R$ from FP calculations of French and Redmer~\cite{FrenchPoP} using both the Perdew, Burke, and Ernzerhof (PBE) and Heyd, Scuseria, and Ernzerhof (HSE) functionals for exchange and correlation. It was anticipated that the HSE functional, which includes the nonlocal Fock exchange, would prove to be more accurate in the calculation of $R$, as this functional has been shown to better reproduce the band gap in semiconductor materials (PBE is known to significantly underestimate the band gap). In comparison with $R$ data of Celliers \textit{et al.}~\cite{Celliers} it would appear that the HSE calculations are less accurate. However, our data suggest a much lower peak $R$, which is in significantly better agreement with the HSE calculations. We note that two recent data points ($\sim$140 and 260 GPa) from a group~\cite{Ozaki} at the Gekko laser in Japan also suggest lower $R$, in very good agreement with our results. These new results lend confidence to the FP calculations, which also predict a super-ionic phase of water at low $T$ and high $\rho$ conditions relevant to planetary interiors. Furthermore, these results strongly suggest that at these conditions water is in a plasma phase, which would imply that a $T$=0 K EOS for water is not sufficient for modeling of hot and mini-Neptunes, and that water would be expected to mix in the H/He envelope rather than form an ice shell separate from an outer H/He envelope.

We presented data with unprecedented accuracy for shock compression of water to 0.7 TPa and 3.8 g/cc in a regime relevant to water-rich models of Uranus, Neptune and the exoplanet GJ436b. The experimental $P$, $\rho$, and $R$ are in excellent agreement with density functional theory predictions, thereby validating first-principles thermodynamic calculations as a sound basis for planetary modeling, and strongly advocating the FP EOS be the standard in modeling water in Neptune, Uranus, and ``hot Neptune'' exoplanets. In particular this work supports the prediction of a $\sim$20\% cooler core temperature for Neptune and Uranus~\cite{Fortney}. As the calculated amount of H and He in the planets decreases with the stiffness of the water EOS, confidence in the presence of a few percent H and He in the deep interior of Neptune and Uranus, as derived from the (rather stiff) FP EOS based models ~\cite{FrenchPRB2009,Fortney}, is strengthened by this work. As H would be metallic, this might influence the generation of the magnetic field. Furthermore, the validation of the FP EOS in the regime relevant to planetary interiors all but eliminates one significant source of uncertainty in the predicted radius evolution of Neptune-mass planets within assumed composition models. This will improve our understanding of the interior structure of these planets, and perhaps our understanding of these planetary systems.

We acknowledge the crew of the Sandia Z facility for their contributions to these experiments, AB and MB for assistance in numerical calculations, and support from the DFG via the SFB 652 and the grant Re 882/11-1. Sandia National Laboratories is a multi-program laboratory managed and operated by Sandia Corporation, a wholly owned subsidiary of Lockheed Martin Corporation, for the U.S. Department of Energy's National Nuclear Security Administration under contract DE-AC04-94AL85000.

%\bibliography{RefsWaterPRL2011}

%\begin{thebibliography}{10}
%\end{thebibliography}

\end{document}